\documentclass{phc-proc4-auth}

\usepackage{amssymb}
\usepackage{epsfig}

\begin{document}

\begin{frontmatter}

\title{Electronic compressibility and charge imbalance relaxation  
in cuprate superconductors}

\author[label1]{Ch. Helm\corauthref{cor1}},
\corauth[cor1]{Corresponding author. Tel.: ++41-(0)1-633 2573; fax: 
++41-(0)1-633 1115; email: helm@phys.ethz.ch}
\author[label2]{L.N. Bulaevskii},
\author[label3]{D.A. Ryndyk},
\author[label3]{J. Keller},
\author[label4]{S. Rother},
\author[label4]{Y. Koval},
\author[label4]{P. M{\"u}ller},
\author[label5]{R. Kleiner}

\address[label1]{Institut f{\"u}r Theoretische Physik,  ETH H{\"o}nggerberg, 
8093 Z{\"u}rich, Switzerland}
\address[label2]{Los Alamos National Laboratory, Los Alamos, NM 87545}
\address[label3]{Institut f{\"u}r Theoretische Physik, Universit{\"a}t 
Regensburg, D-93040 Regensburg, Germany} 
\address[label4]{Physikalisches Institut III, 
Universit{\"a}t Erlangen-N{\"u}rnberg,D-91058 Erlangen,  Germany}
\address[label5]{Physikalisches Institut, Universit{\"a}t T{\"u}bingen, 
D-72076 T{\"u}bingen,Germany}

\begin{abstract}
In the material SmLa$_{1-x}$Sr$_x$CuO$_{4-\delta}$ with 
alternating intrinsic Josephson junctions we  explain theoretically the relative
amplitude of the two plasma peaks in transmission by taking into
account the spatial dispersion of the Josephson Plasma Resonance in 
$c$ direction due to charge coupling. From this and the magnetic field dependence 
of the plasma peaks in the vortex solid and liquid states it is shown that the 
electronic compressibility of the CuO$_2$ layers is consistent with a 
free electron value. Also the London penetration depth $\lambda_{ab} \approx 
1100 {\rm \AA}$ near $T_c$ can be determined. The voltage response in the $IV$-curve of 
a Bi$_2$Sr$_2$CaCu$_2$O$_8$ mesa due to microwave irradiation or current 
injection in a second mesa is related to the  nonequilibrium charge imbalance of 
quasiparticles and Cooper pairs and from our experimental data the relaxation
time $\sim 100 {\rm ps}$ is obtained. 
\end{abstract}


\end{frontmatter}

The theoretical analysis of optical and transport
properties of intrinsic Josephson junctions in high-$T_c$ superconductors
allows to extract key microscopic parameters of the CuO$_2$-layers, such as
the electronic compressibility and the charge imbalance relaxation rate, which 
are hard to obtain otherwise.

\begin{figure}
\epsfig{file=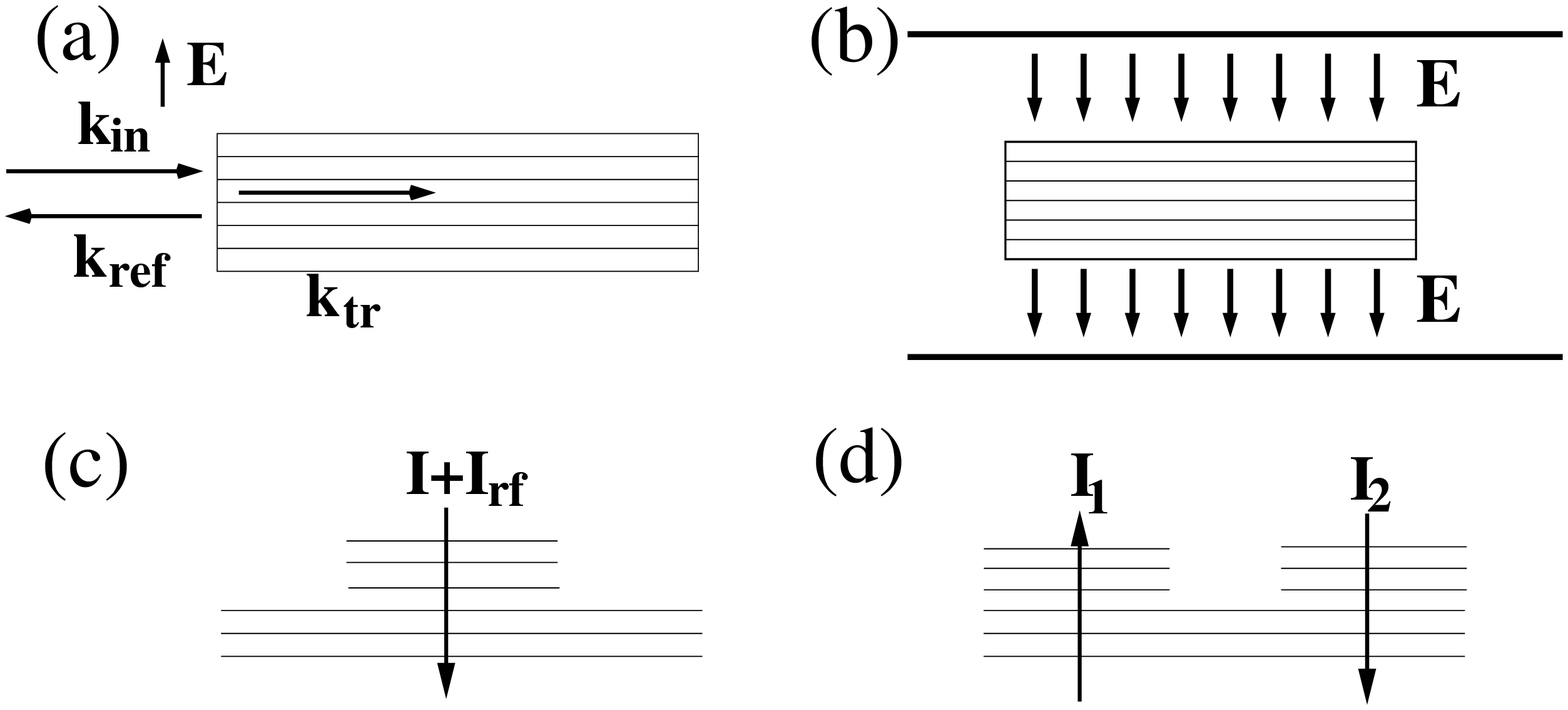,width=0.45\textwidth,clip=}
\caption{Geometry for measuring (a) the reflectivity with parallel
incidence, (b) the microwave absorption in a cavity, (c) Shapiro steps, and 
(d) dc-voltage response on dc-current injection}
\label{geometry}
\end{figure}
\begin{figure}
\epsfig{file=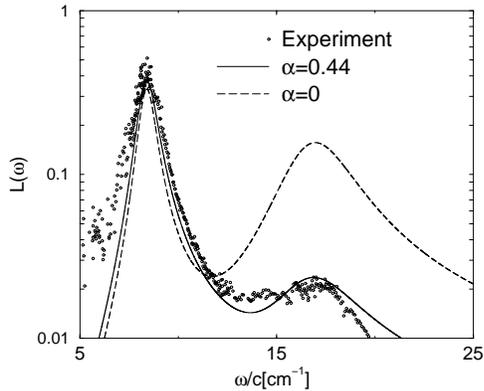,width=0.33\textwidth,angle=-90,clip=}
\caption{
Experimental data [3] together with our calculated loss function $L(\omega)$
for $\sigma_1 = r \sigma_2=0.12 \epsilon_0 \omega_{0,1} / 4 \pi$, 
$r= \omega_{0,1}^2/ \omega_{0,2}^2$ 
(solid: $\alpha=0.44$, $r=0.38$, $\omega_{0,1}/ c=7.2~{\rm cm}^{-1}$, 
$\epsilon_{0}=18$; 
dashed: $\alpha=0$, $r=0.24$, $\omega_{0,1}/c=8.3~{\rm cm}^{-1}$, $\epsilon_{0}=13$). 
\label{losspic}
}
\end{figure}
In optical transmission or reflectivity experiments (cf. Fig.~\ref{geometry}a) 
the Josephson Plasma 
Resonance (JPR) of Cooper pairs oscillating between the layers, 
creates a peak in the loss function $L(\omega)$ (e.g. in TBCCO \cite{verner}). 
Recently, in the novel material SmLa$_{1-x}${}Sr$_x$CuO$_{4-\delta}$ with
alternating junctions two plasma peaks were observed with a high ratio of their  
amplitudes \cite{smlaexp,gorshunov}. This can only be explained, if the
spatial dispersion of the JPR in $c$ direction due to charge fluctuations of 
the superconducting condensate, is taken into account, see Fig.~\ref{losspic}
and Ref. \cite{prl}. Using the 
data in \cite{smlaexp,gorshunov} we find that 
the electronic compressibility of the layers is consistent with a free
electron value. This corresponds to a $c$ axis 
dispersion $\sim \alpha$ of the (bare) JPR frequencies, 
$\omega_{l}^2 = \omega_{0,l}^2 (1 + \alpha k_z^2)$, $l=1,2$
 with  $\alpha = (\epsilon_0 / 4 \pi e s )  (\partial \mu / \partial \rho)
\approx 0.4$ ($\epsilon_0$ background dielectric constant, $s=6.3 {\rm \AA}$ 
interlayer distance, $\mu$ chemical potential, $\rho$ 2d charge density). 
This result is confirmed by the magnetic field dependence of the plasma peaks 
\cite{pim}, using the dependence of the critical current densities, which 
is well known in the vortex liquid state and has been newly derived for 
the vortex solid 
\cite{vortexsolid}. The parameter $\alpha$ is not only relevant to understand the 
coupled dynamics of stacks of intrinsic Josephson junctions, e.g. for 
THz applications, but might also provide an
important input for microscopic theories of high-$T_c$-superconductivity. 
Beyond that, the $c$-axis quasiparticle conductivity 
$\sigma_1 \sim 4-10({\rm \Omega m})^{-1}$ and the
London penentration depth $\lambda_{ab} =1100 {\rm \AA}$ at $T=10 {\rm K}$
(near $T_c$) parallel
to the layers can be accurately determined. 
Our theory for $L(\omega)$ quantitatively improves Ref.~\cite{marel4} by 
taking into account the different conductivities $\sigma_l$ in 
the junctions in accordance with their different current densities and the
influence of the discrete atomic structure. Our results are in principle 
also applicable for microwave absorption experiments in a cavity 
(Fig.~\ref{geometry}b). 

From a theoretical point of view the inclusion of the wavevector dependence of 
the dielectric function $\epsilon (\omega, {\bf k})$ poses a nontrivial 
problem and requires to go beyond the conventional Fresnel theory \cite{epl}. 
Generally, for systems with spatial dispersion (e.g. phonons 
\cite{ivo}), multiple eigenmodes with different group velocities are excited 
in the crystal and at certain extremal frequencies,
where the group velocity vanishes, the atomic structure of the crystal enters
explicitly in optical properties. The possibility to stop light dynamically
by affecting the JPR in an external magnetic field might serve as a 
building block for a future magneto-optical device, e.g. to store
photonic qubits \cite{epl}.

Further, we perform two different transport experiments 
on Bi$_2$Sr$_2$CaCu$_2$O$_8$, in order to investigate nonequilibrium 
effects beyond the coupling $\alpha$ \cite{rother}: 
(1) In $2$-point measurements of the $IV$-curves
in the presence of high-frequency irradiation ($I_{\rm rf}$) 
of frequency $f$ a shift of the 
voltage of Shapiro steps of $\sim 3\%$ from the canonical value $V_s = \hbar f / 2 e$
has been observed due to the resistance at the $NS$-contact 
(Fig.~\ref{geometry}c, first discussed in \cite{sendai}). 
(2) In the $IV$-curves of double-mesa structures an influence
of the dc voltage $V_1$ measured at one mesa on the dc quasiparticle current 
$I_2$ injected 
into the other mesa is detected (Fig.~\ref{geometry}d).  
Both effects can be explained by charge-imbalance on the superconducting layers 
between resistive and superconducting junctions, where Cooper pairs and 
quasiparticles are transformed into each other.  
With the help of a recently developed theory \cite{dimatheory} we get
the charge imbalance relaxation time as $\approx 70-450 {\rm ps}$ 
(depending on the sample).

In conclusion, the optical spectroscopy of the Josephson 
plasma resonance in SmLa$_{1-x}$Sr$_x$CuO$_{4-\delta}$ suggests a free 
electronic compressibility of the superconducting CuO$_2$-layers and a 
London penetration depth $\lambda_{ab}=1100 {\rm \AA}$ near $T_c$. From the
analysis of the resistive state in the presence of microwave
irradiation or current injection a charge imbalance relaxation time of the
order $\sim 100 {\rm ps}$ is obtained.  

The authors thank for fruitful discussions with G. Blatter, M. Dressel, 
B. Gorshunov, D. van der Marel, A. Pimenov, and Z.-X. Shen and acknowledge financial 
support of the Swiss NCCR "MaNEP", the U.S. DOE, the German DFG and the 
Bayerische Forschungsstiftung.


\begin{thebibliography}{00}

\bibitem{verner} V. K. Thorsm$\o$lle et al., 
%R. D. Averitt, M. P. Maley, L. N. Bulaevskii, C. Helm, A. J. Taylor, 
Optics Letters 26 (2001) 1292; Physica B 312 (2002) 84.

\bibitem{smlaexp}D. Duli\'{c} et al.,
% A. Pimenov, D. van der Marel, D.M. Broun, S. Kamal, 
%W.N. Hardy, A.A. Tsvetkov, I.M. Sutjaha, R. Liang, A.A. Menovsky, A. Loidl, 
%and S.S. Saxena, 
Phys.Rev.Lett. 86 (2001) 4144;  H. Shibata, Phys.Rev.Lett. 86 (2001) 2122; 
T. Kakeshita et al., 
%S. Uchida, K.M. Kojima, S. Adachi, S. Tajima, B. Gorshunov, M. Dressel, 
Phys.Rev.Lett. 86 (2001) 4140. 

\bibitem{gorshunov} B. Gorshunov et al., to be published. 

\bibitem{prl} Ch.  Helm, L.N. Bulaevskii, E.M. Chudnovsky,  M.P. Maley, 
Phys.Rev.Lett. 89  (2002) 057003.

\bibitem{pim}A. Pimenov et al., 
%A. Loidl, D. Duli\'{c}, D. van der Marel, I.M.  Sutjahja, A.A. Menovsky, 
Phys.Rev.Lett. 87 (2001) 7003. 

\bibitem{vortexsolid} L.N. Bulaevskii, C. Helm, Phys.Rev.B 66  (2002) 174505. 

\bibitem{marel4} D. van der Marel et al., Phys.Rev.B 64 (2001) 24530.

\bibitem{epl} L.N. Bulaevskii, Ch. Helm, A.R. Bishop, M.P. Maley, 
Europhys. Lett. 58  (2002) 415; Ch. Helm, L.N. Bulaevskii, Phys.Rev.B 66  
(2002) 094514. 

\bibitem{ivo} I. Kaelin, C. Helm, G. Blatter, physics/0303046.

\bibitem{rother} S. Rother, Y. Koval, P. M\"uller, R. Kleiner, D.A. Ryndyk, 
J. Keller, Ch. Helm, Phys.Rev.B 67 (2003) 024510. 

\bibitem{sendai} 
C. Helm et al.,  
%J. Keller, C. Preis, A. Sergeev,
 Physica C 362 (2001) 43. 

\bibitem{dimatheory} 
D. A. Ryndyk, J. Keller, C. Helm, J.Phys.: Cond. Mat. 14 (2002) 815. 

\end{thebibliography}
\end{document}